\documentclass[%
 aip,
 amsmath,amssymb,
preprint,%
]{revtex4-2}

\usepackage{graphicx}
\usepackage{dcolumn}
\usepackage{bm}
\usepackage{mathrsfs}
\usepackage{subcaption}
\usepackage[utf8]{inputenc}
\usepackage[T1]{fontenc}
\usepackage{mathptmx}
\usepackage{etoolbox}
\usepackage{hyperref}
\usepackage[capitalize]{cleveref}

\makeatletter
\def\@email#1#2{%
 \endgroup
 \patchcmd{\titleblock@produce}
  {\frontmatter@RRAPformat}
  {\frontmatter@RRAPformat{\produce@RRAP{*#1\href{mailto:#2}{#2}}}\frontmatter@RRAPformat}
  {}{}
}%
\makeatother
\begin{document}


	\title{Photonic frequency multiplexed next-generation reservoir computer}

\author{Nicholas Cox}
\email{nicholas.a.cox34.civ@us.navy.mil} 
\author{Joseph Murray}
\author{Joseph Hart}
\author{Brandon Redding}

\affiliation{U.S. Naval Research Laboratory, 4555 Overlook Ave, SW, Washington, DC 20375, USA}

\date{\today}

\begin{abstract}

In this work, we introduce and experimentally demonstrate a photonic frequency-multiplexed next generation reservoir computer (FM-NGRC) capable of performing real-time inference at GHz speed. NGRCs apply a feed-forward architecture to produce a feature vector directly from the input data over a fixed number of time steps. This feature vector, analogous to the reservoir state in a conventional RC, is used to perform inference by applying a decision layer trained by linear regression. Photonic NGRC provides a flexible platform for real-time inference by forgoing the need for explicit feedback loops inherent to a physical reservoir. The FM-NGRC introduced here defines the memory structure using an optical frequency comb and dispersive fiber while the sinusoidal response of electro-optic Mach-Zehnder interferometers controls the nonlinear transform applied to elements of the feature vector. A programmable waveshaper modulates each comb tooth independently to apply the trained decision layer weights in the analog domain. We apply the FM-NGRC to solve the benchmark nonlinear channel equalization task; after theoretically determining feature vectors that enable high-accuracy distortion compensation, we construct an FM-NGRC that generates these vectors to experimentally demonstrate real-time channel equalization at 5 GS/s with a symbol error rate of $\sim 2\times 10^{-3}$. 

\end{abstract}

\maketitle

\section{Introduction}\label{sec:introduction}
Recent years have seen dramatic growth in the use of artificial neural networks for research and commercial purposes. Much of this progress has been achieved by improving the industry-standard technology of deep neural networks implemented on graphical processing units (GPUs) \cite{lecunDeepLearning2015,alzubaidiReviewDeepLearning2021}.  
While massive parallelization combined with techniques such as batch processing have driven remarkable growth in the capability of deep neural networks, the inference rate and latency in GPU-based systems remain constrained by digital clock rates. This clock rate limitation makes it challenging to apply conventional GPU-based neural networks in applications requiring continuous, real-time operation on data at $>$GHz rates, such as high-bandwidth optical communications and RF signal processing. 

Photonic neural networks, which exploit the large bandwidth and rapid modulation capacity of optical signals, have the potential to overcome this limitation and enable real-time inference at GHz data rates \cite{wetzsteinInferenceArtificialIntelligence2020}. Photonic reservoir computers (RCs), which specialize in processing time-dependent data, are particularly well-suited for these types of applications \cite{jaegerHarnessingNonlinearityPredicting2004,maassRealTimeComputingStable2002}. RCs operate by feeding data into a fixed randomly-connected nonlinear feedback loop and training a single readout layer to map the optical signal within the reservoir (the reservoir state) to a desired output via linear regression. The use of a fixed, un-trained feedback loop as the reservoir has enabled a wide variety of photonic implementations, including systems utilizing temporal \cite{appeltantInformationProcessingUsing2011,paquotOptoelectronicReservoirComputing2012,martinenghiPhotonicNonlinearTransient2012,brunnerParallelPhotonicInformation2013,hickeInformationProcessingUsing2013,dejonckheereAllopticalReservoirComputer2014,vinckierHighperformancePhotonicReservoir2015,duportVirtualizationPhotonicReservoir2016,keuninckxRealtimeAudioProcessing2017,largerHighSpeedPhotonicReservoir2017,nguimdoPredictionPerformanceReservoir2017,ortinReservoirComputingEnsemble2017,argyrisPhotonicMachineLearning2018,takanoCompactReservoirComputing2018,vatinExperimentalReservoirComputing2019,argyrisPAM4Transmission15502019,tsurugayaCrossgainModulationbasedPhotonic2022,huangMultitaskPhotonicTimedelay2022,jacobsonHybridConvolutionalOptoelectronic2022,fangBidynamicalAllopticalReservoir2023,guoExperimentalDemonstrationModulation2023,piccoHighSpeedHuman2023,shenDeepPhotonicReservoir2023,tangAsynchronousPhotonicTimedelay2023,pengCoherentAllopticalReservoir2024,henaffOpticalPhaseEncoding2024,guoShorttermPredictionChaotic2024,renPhotonicTimedelayedReservoir2024,zhouPhotonicDeepResidual2024}, spatial \cite{vandoorneExperimentalDemonstrationReservoir2014,nakajimaScalableReservoirComputing2021,sackesynExperimentalRealizationIntegrated2021,boikovScalableDelayLineFree2023,gooskensExperimentalResultsNonlinear2023,maIntegratedPhotonicReservoir2023,masaadOptoelectronicMachineLearning2024}, and frequency multiplexing \cite{butschekPhotonicReservoirComputer2022,lupoDeepPhotonicReservoir2023,liScalableWavelengthmultiplexingPhotonic2023} to encode reservoir nodes. Achieving ultra-low latency is particularly challenging for time-multiplexed architectures, for which the decision layer for an RC with $N$ nodes requires modulating the virtual nodes $N$ times faster than the desired inference rate \cite{smerieriAnalogReadoutOptical2012,duportFullyAnaloguePhotonic2016}. While spatial and frequency-multiplexed RCs avoid this constraint, their speeds are still limited by the round-trip time of the photonic reservoir. Most RCs capable of high-speed operation reduce this round-trip time using passive photonic integrated circuit (PIC)-based cavities. While these PIC-based designs have enabled impressive performance, including 20 GHz inference rates \cite{maIntegratedPhotonicReservoir2023}, their limited control over the memory structure and nonlinear transform make it challenging to optimize them for different problem types. 

A newer class of RC forgoes the reservoir feedback in favor of a completely feed-forward design. This design is referred to as either feed-forward RC or next-generation RC (NGRC), following the work of \textcite{gauthierNextGenerationReservoir2021} which demonstrated comparable performance to standard RC. NGRC operation relies on forming a feature vector by applying a series of linear and nonlinear transformations to the input data at a fixed set of consecutive delays. This feature vector replaces the "reservoir state" of a conventional RC and a single trained readout layer is used to map the feature vector to the desired output. To date, photonic NGRCs have been implemented with temporal multiplexing and coherent node mixing via Rayleigh backscattering \cite{coxPhotonicNextgenerationReservoir2024} and spatial multiplexing with coherent node mixing by scattering in a disordered medium \cite{wangOpticalNextGeneration2024} or on-chip multimode waveguides \cite{wang103TOPSMm$^2$Integrated2024}. 
However, none of these designs have performed inference in real-time, relying instead on digitizing the feature vector and applying the decision layer in software. In addition, these initial photonic NGRCs formed feature vectors by a random nonlinear projection of the input data over a fixed number of time steps. While this scheme works well for a variety of data analysis tasks, it is unlikely to be the most efficient design because many nodes provide redundant information and the nonlinear transformation cannot be optimized for the task at hand. This random-mixing approach does not take advantage of a major strength of NGRC, which is the ability to tailor the feature vector for a given problem in a readily-interpretable manner \cite{gauthierNextGenerationReservoir2021}. 

In this work, we introduce a photonic frequency multiplexed NGRC (FM-NGRC) capable of real-time inference at GHz data rates using an analog decision layer. We also demonstrate how the FM-NGRC feature vector can be tailored to match the requirements of a given task. The FM-NGRC operates by encoding input data onto lines of an optical frequency comb and passively generating consecutive time delays using optical dispersion. Data in the RF electronic domain is encoded onto optical carriers using Mach-Zehnder electro-optic modulators (EOMs), exploiting the inherent sinusoidal modulator response function to generate polynomial-like nonlinear transformations of the input data. By varying the EOM bias points, we can adjust the nature of the nonlinear transform to match the needs of a specific task. The memory is controlled by adjusting the number of frequency comb teeth and the amount of optical dispersion. Crucially, the decision layer can be implemented in the analog photonic domain using a programmable waveshaper (WS) to apply separate weights to each comb tooth. As an initial demonstration, we apply the FM-NGRC to the benchmark channel equalization task \cite{jaegerHarnessingNonlinearityPredicting2004,duportAllopticalReservoirComputing2012,paquotOptoelectronicReservoirComputing2012,vinckierHighperformancePhotonicReservoir2015,duportFullyAnaloguePhotonic2016,duportVirtualizationPhotonicReservoir2016,antonikOnlineTrainingOptoElectronic2017,vandersandeAdvancesPhotonicReservoir2017,takanoCompactReservoirComputing2018,vatinExperimentalReservoirComputing2019,chemboMachineLearningBased2020,butschekPhotonicReservoirComputer2022,linDeepTimeDelayReservoir2023,lupoDeepPhotonicReservoir2023,xuPhotonicReservoirComputing2023,zhouPhotonicDeepResidual2024}. This task requires the FM-NGRC to undo the effects of a channel with intersymbol interference, nonlinear distortion and additive Gaussian noise. The symbol error rate (SER) of the equalized data serves as the performance metric. We first provide a theoretical basis for the system operation, deriving the form of a feature vector suitable for the channel equalization problem. We then experimentally construct an FM-NGRC following this design, demonstrating real-time channel equalization at 5 GS/s with a symbol error rate of $\sim2\times10^{-3}$. 

\section{Experimental design} \label{sec:experiment}
This section provides a high-level description of the NGRC architecture followed by a description of the photonic implementation. The NGRC begins by generating a feature vector directly from a series of $m$ consecutive time samples. In general, the NGRC feature vector can contain an arbitrary combination of nonlinear transformations and interaction terms between the $m$ time samples provided at a given time-step. The decision layer maps this feature vector to the desired result by computing the dot product with a vector of trained output weights. As in conventional RC, only these output weights are explicitly trained. The form of the feature vector is determined through hyper-parameter optimization, demonstrating the ability to tailor the NGRC to the task at hand in an intrepretable fashion.  

\begin{figure}[h]
	\centering
	\includegraphics[scale=1.0]{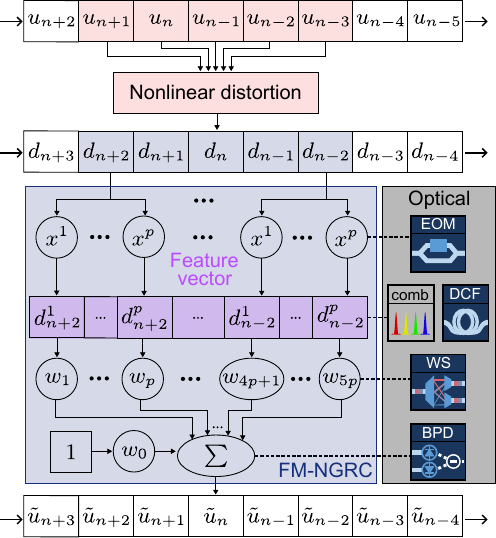}
	\caption{A block diagram of the operation performed by the FM-NGRC to reverse the effects of nonlinear distortion. Circled elements represent mathematical operations, including monomial transformation ($x^i$), weight multipliciation ($w_i$), and summation ($\sum$). The right-hand side shows the optical components that implemented each step. Electro-optic modulators (EOMs) applied nonlinear transforms to input data, a frequency comb in conjunction with dispersion compensating fiber (DCF) provided temporal overlap between adjacent time samples to form a feature vector, and a waveshaper (WS) combined with a balanced photodetector (BPD) applied the readout weights to perform inference.}
	\label{fig:block}
\end{figure}

In this work, we focus on the problem of distortion compensation, or channel equalization, which is particularly relevant for high-bandwidth fiber optic communication in which optical nonlinearites in the fiber distort the original data. In this problem, an undistorted data sequence $u_n$ (where $n$ denotes the time step) passes continuously through a nonlinear filtering process which induces intersymbol interference (ISI) between a fixed number of time samples. The output of this process is a series of distorted values $d_n$. The NGRC, depicted in \cref{fig:block}, is designed to reverse this distortion process and recover the original data $u_n$. At time step $n$, an input vector is formed by collecting $m$ adjacent time samples of the distorted data series (shown as the highlighted elements in the $d_n$). While the illustration depicts a length $m=5$ input vector centered at $d_n$, the number of consecutive samples before and after time $n$ can be adjusted independently based on the memory required for a given task. The system then applies element-wise monomial transformations of order $0$ to $p$, represented in \cref{fig:block} by circles labeled with the transform order. The result is grouped into an output feature vector $\mathbf{v}_{\text{out},n}$. The feature vector is mapped to a desired output by multiplying weight values (represented by encircled elements $w_i$) and summing the result. The optimal weights, $w_i$, required to reconstruct the original data point $d_n$ are found through linear regression using labelled training data. After the training process, the system behaves as an approximation to the inverse of the nonlinear filter imposed by the channel. Note that a more general feature vector could also include interaction terms between elements in the feature vector shown in \cref{fig:block}\cite{gauthierNextGenerationReservoir2021}; however, as we will show, these interaction terms are not required for the channel equalization task \cite{jaegerHarnessingNonlinearityPredicting2004} examined here.

\begin{figure}[!h]
	\centering
	\includegraphics[scale=1]{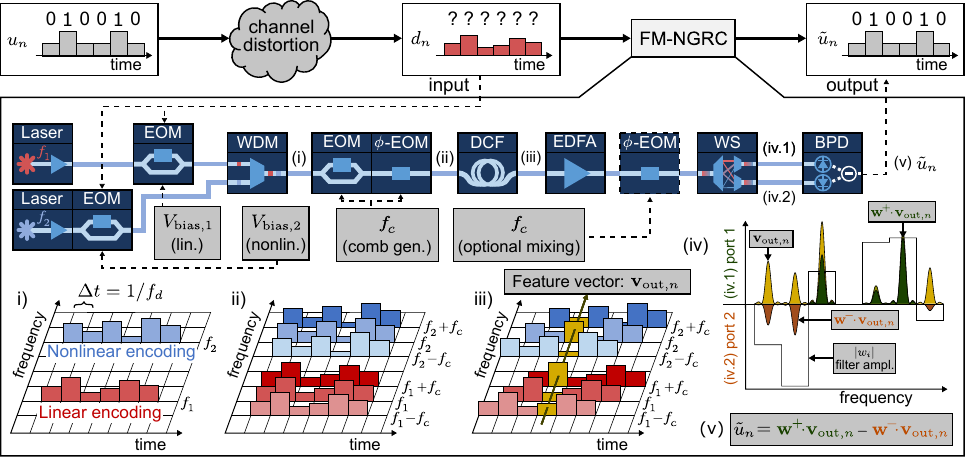}
	\caption{A diagram of the FM-NGRC apparatus used to reverse the effects of distortion by a nonlinear channel. EOM: electro-optic modulator, WDM: wavelength division multiplexer, EDFA: Erbium-doped fiber amplifier, DCF: dispersion compensating fiber, WS: Waveshaper, BPD: balanced photodetector.}
	\label{fig:experiment}
\end{figure}

The photonic FM-NGRC was designed to implement the process shown in \cref{fig:block} in real time at GHz data rates. This experimental FM-NGRC operated in four functional blocks:  (1) values of $d_n$ were encoded onto multiple optical carriers, exploiting the nonlinear Mach-Zehnder EOM response to generate the different element-wise monomial transforms shown in \cref{fig:block}), (2) The feature vector was formed by generating an optical frequency comb centered around each carrier and using dispersion compensating fiber (DCF) to temporally align $m$ sequential elements of $d_n$, (3) a waveshaper applied the output weights by scaling the contribution of each comb tooth, and (4) a balanced photodetector summed the weighted feature vector to yield the recovered signal. A schematic diagram showing the FM-NGRC architecture is shown in \Cref{fig:experiment}. 

The photonic FM-NGRC implemented in this work employed two element-wise monomial transforms on $m=9$ sequential samples of $d_n$ to generate a feature vector with 18 elements. The transformations were performed by encoding the distorted data $d_n$ on two laser sources at different frequencies (193.2 THz and 194.4 THz) using two Mach-Zehnder interferometer (MZI)-based EOMs with different bias points. The first EOM was biased at quadrature to provide linear encoding while the second EOM was biased at the minimum transmission (null bias) point to provide quadratic encoding. We also experimentally optimized the bias point on the second EOM, finding that biasing between quadrature and minimum transmission provided more robustness to noise. The data was encoded at 5 GHz using a single arbitrary waveform generator and directed to the two EOMs. In the case of optical communication, this encoded voltage could come directly from an optical signal using a photodetector. The RF power reaching the linear EOM was limited to $V_{\pi}/8$ to ensure a primarily linear response, while the RF power applied to the second EOM was amplified to $~V_{\pi}/3$ to increase the optical signal level and the magnitude of higher-order nonlinearities generated when biasing at the optimal point. 

The optical signals containing a linear and nonlinear copy of the input data were then combined using a wavelength division multiplexer (WDM), as shown in inset (i) in \Cref{fig:experiment}. We then generated $m=9$ copies of the linear and nonlinear encoded data on an optical frequency comb. The comb was generated by driving EOMs with a sinusoudal voltage at a rate of  $f_c = 15.411$ GHz, which sets the comb tooth spacing. This comb spacing was chosen so that, upon passing through dispersion compensating fiber, adjacent lines are delayed by exactly the duration of a single data point encoded at a 5 GHz rate. In order to generate 9 comb teeth with comparable amplitude, we used a MZI-based intensity EOM followed by two phase-only EOMs (for simplicity, \cref{fig:experiment} represents the phase-only EOMs as a single $\phi$-EOM block). By adjusting the bias point of the MZI-based EOM, we were able to generate 9 comb teeth with less than 10 dB of variation in power. Inset (ii) in \cref{fig:experiment} shows a simplified example with 3 comb teeth.  

The frequency combs were then directed through 100 km of dispersion compensating fiber, introducing a total group delay dispersion of $\text{GDD} = 200 \text{ ps}/(15.411 \text{ GHz})$. This group delay ensures that, at a given time step $n$, the optical pulse contains 9 consecutive time samples centered around the source frequency $f_s$ of each laser, i.e. $f_{s,k} = f_s + k f_c$ for $k = -4,\dots,4$ and $s = 1,2$. To compensate for the different dispersion experienced by the combs centered around the two laser sources, the RF signal driving the linear-encoding EOM was delayed using an RF phase shifter. As shown in inset (iii) in \cref{fig:experiment}, this produced a feature vector at time step $n$ of length $2m$, comprising the linear and nonlinear encoded elements in the input data over $m$ time steps. At this point, the feature vector could be modified to include interaction terms (e.g., $d_nd_{n-1}$) using a $\phi$-EOM, driven either at $f_c$ or some integer division to generate additional comb lines between the original comb tooth frequencies. However, we found that these additional terms were not required to solve the benchmark channel eqalization task and did not include the additional $\phi$-EOM stage in this initial demonstration.  

After training, weights were applied to the feature vector to recover the original input data, $u_n$. In the photonic FM-NGRC, weights were introduced using an optical waveshaper (WS, Finisar WaveShaper 4000a) which applied varying levels of attenuation to each comb tooth. Positive and negative weights were accomodated by directing comb teeth with positive weights to the first output port of the waveshaper and teeth with negative weights to the second output port. As shown in inset (iv) in \cref{fig:experiment}, the weighted frequency comb was then directed to a balanced photodetector (PD). The output of the balanced detector, $\widetilde{u}_n$, should approximate the original undistorted data, $u_n$, as shown in inset (v) in \cref{fig:experiment}. Based on the equipment at hand, we implemented balanced detection using two separate 10 GHz photodetectors (PDs) and subtracted the recorded photovoltages. To test the FM-NGRC performance, we implemented a simple receiver using an oscilloscope to digitize photovoltages at a $100$ GS/s rate with a 10-point moving average filter. In software, the signal was down-sampled to a rate of 5 GS/s to match the input encoding rate and then thresholded to assign a value for the reconstructed data point.

In the training phase, we needed to measure the feature vector produced by labelled training data. This required us to record the amplitude of each comb tooth, and was accomplished by using the waveshaper as a programmable filter to transmit a single comb tooth at time to the first output port (iv.1 in \cref{fig:experiment}). The training data was repeated 18 times, allowing us to record each comb tooth in sequence. For the final demonstration using analog output weights, we repeated this process for each PD to account for slight differences in the response of the two PDs. A detailed description of the training and inference procedures for digital and analog weighting experiments are described in \cref{sec:digital} and \cref{sec:analog}, respectively.

\section{Results}

We used the FM-NGRC platform to solve the nonlinear channel equalization task first proposed by \textcite{jaegerHarnessingNonlinearityPredicting2004}. Beyond acting as a benchmark to compare RC performance, this equalization problem is a simplified version of the nonlinear compensation required for high bandwidth communications signals transported through optical fiber \cite{argyrisPhotonicMachineLearning2018,argyrisPAM4Transmission15502019,katumbaNeuromorphicSiliconPhotonics2019,ranziniTunableOptoelectronicChromatic2019,darosReservoirComputingBasedEqualization2020,argyrisComparisonPhotonicReservoir2020,sackesynExperimentalRealizationIntegrated2021,pengPhotonicsInspiredCompactNetwork2022,luSignalRecoveryOptical2022,liScalableWavelengthmultiplexingPhotonic2023,shenDeepPhotonicReservoir2023,gooskensExperimentalResultsNonlinear2023,zuoIntegratedSiliconPhotonic2023,huangSiliconPhotonicElectronic2021,wangMultiWavelengthPhotonicNeuromorphic2023,pengCoherentAllopticalReservoir2024}. 
Studying the simplified equalization problem also allows us to make sense of the FM-NGRC operation through analytical examination and simulations in \cref{sec:eq,sec:sim}, the results of which are corroborated in the experimental data presented in \cref{sec:digital,sec:analog}.

\subsection{Nonlinear channel equalization}
\label{sec:eq}

The simplified nonlinear channel equalization problem supposes an undistorted communications signal is formed by a vector of elements $u_n$ ($n = 1,\dots, N_{\text{total}}$) drawn from the independent and identically distributed random variable $S = \{-3,-1,1,3\}$. This information passes through a channel characterized by intersymbol interference, element-wise nonlinear transformation, and additive Gaussian noise. The distortion process begins with a linear convolution of the input data, generating an intermediate variable,
\begin{align}\label{eq:qn}
	q_n &= 0.08 u_{n+2} - 0.12 u_{n+1} + u_{n} + 0.18 u_{n-1} - 0.1 u_{n-1} \nonumber \\
	 &+ 0.091 u_{n-3} - 0.05 u_{n-4} + 0.04 u_{n-5} + 0.03 u_{n-6} + 0.01 u_{n-7},
\end{align}
followed by a nonlinear transformation and the addition of noise
\begin{equation}\label{eq:dn}
	d_n = q_n + 0.036 q_n^2 - 0.01 q_n^3 + \nu_n.
\end{equation}
The noise contribution ($\nu_n$) is sampled from a Gaussian distribution with zero mean and variance $\sigma^2 = \sigma_{u}^2\cdot10^{\text{SNR (dB)}/10}$, where $\sigma_u^2 = 5$ is the variance of the undistorted sequence, $u_n$. Here, we will investigate the SER for simulated channel SNR values ranging from 12 to 48 dB.

The FM-NGRC task is to reconstruct the original data, $u_n$, from the distorted sequence, $d_n$. Examining \cref{eq:qn,eq:dn}, this system can be inverted in two steps. First, we aim to find a nonlinear transform $f_{\text{NL}}$ such that $f_{\text{NL}}(d_n) = q_n$. Because \cref{eq:dn} is monotonic throughout most of the range of interest, this inverse can be approximated using polynomials. Once the nonlinearity is corrected, we must invert the linear convolution by finding the filter $f_{\text{L}}$ such that $(f_{\text{L}} * q)_n = d_n$. Based on this high-level analysis, we expect that an NGRC that generates a feature vector containing element-wise monomial terms over $m\sim9$ delays (based on the length of the convolution in \cref{eq:qn}) should be able to recover the original data, $u_n$.

\subsection{Simulation and principle of operation}
\label{sec:sim}
This section illustrates the operating principles of the FM-NGRC through analytical and numerical simulation. Recall that \cref{sec:eq} introduced the idea that the channel equalization should perform two sub-tasks: we must first invert the nonlinear transform to convert $d_n$ to $q_n$ and then find the linear kernel which convolves with the reconstructed $q_n$ to yield $\tilde{u}_n$, an approximation to the undistorted input data $u_n$. 

For the region in which the function $d(q)$ is monotonic, the inverse function $q(d)$ is single-valued. Therefore, there exists some order-$p$ polynomial expansion such that
\begin{equation}\label{eq:inv}
	q_n \approx a_0 + \sum_{i = 1}^p a_i d_n^i
\end{equation}
for the element at time step $n$. In this work, we will show that $p=2$ suffices to yield a satisfactory symbol error rate for the equalization of the channel in \cref{eq:qn,eq:dn}. To illustrate how this inverse function is combined with the deconvolution function in the NGRC, consider a simplified feature vector $\mathbf{v}_{\text{out},n}$ with $p=2$ and $m=3$ time-delay elements, combined with a unity scalar element:
\begin{equation}\label{eq:vout_analytic}
\mathbf{v}_{\text{out},n} = 
	\begin{bmatrix}
		 1 & d_n & d_n^2 & d_{n - 1} & d_{n - 1}^2 & d_{n - 2} & d_{n-2}^2
	\end{bmatrix}.
\end{equation}
We can then define an inversion operator $A$ that acts on the vector $\mathbf{v}_{\text{out},n}$ to yield the vector
\begin{equation}
	\mathbf{q}_n = A \mathbf{v}_{\text{out},n}^T  = \begin{bmatrix}
		a_0 & a_1 & a_2 & 0 & 0 & 0 & 0 \\
		a_0 & 0 & 0 & a_1 & a_2 & 0 & 0 \\
		a_0 & 0 & 0 & 0 & 0 & a_1 & a_2 
	\end{bmatrix}
	\begin{bmatrix}
		 1 & d_n & d_n^2 & d_{n - 1} & d_{n - 1}^2 & d_{n - 2} & d_{n-2}^2
	\end{bmatrix}^T,
\end{equation}
where elements $a_i$ come from \cref{eq:inv}. The values $q_{n,i}$ for $i = 1,2,3$ represent consecutive time samples to which we apply a convolution with some unknown kernel $B$, expressed as
\begin{equation}\label{eq:Bq}
	\widetilde{u}_n = \sum_{i = 1}^{m } B_i q_{n,i},
\end{equation}
to approximate the undistorted data point, $u_n$. Re-writing \cref{eq:Bq} in matrix form, we may combine the actions of transformations $A$ and $B$ to find
\begin{equation}\label{eq:unB}
	\widetilde{u}_n = B \mathbf{q}_{n} = BA \mathbf{v}_{\text{out},n}^T = W\mathbf{v}_{\text{out},n}^T.
\end{equation}
\Cref{eq:unB} indicates that determination of the row vector $W$ allows us to approximate the value of $u_n$ from a given feature vector $\mathbf{v}_{\text{out},n}$. In practice, we determine $W$ by performing $N_{\text{train}}$ trials for which $u_n$ and $\mathbf{v}_{\text{out},n}$ are known. The trial results are concatenated into the matrix relation
\begin{equation}\label{eq:WX}
	\begin{bmatrix}
		u_1 & u_2 &\dots & u_{N_{\text{train}}}
	\end{bmatrix} = W \begin{bmatrix}
	\mathbf{v}_{\text{out},1}^T & \mathbf{v}_{\text{out},2}^T & \dots & \mathbf{v}_{\text{out},N_{\text{train}}}^T  
	\end{bmatrix},
\end{equation}
to which we can apply regularized least squares regression to determine the best approximation to $W$. With $W$ known, we apply \cref{eq:unB} in the testing phase to compute $\widetilde{u}_n$. The polynomial order $p$, memory length $m$, and number of memory elements before and after $d_n$ in $\mathbf{v}_{\text{out},n}$ serve as hyperparameters to be adjusted until the best approximation is found.

\begin{figure}[!h]
	\centering
	\includegraphics[scale=1.0]{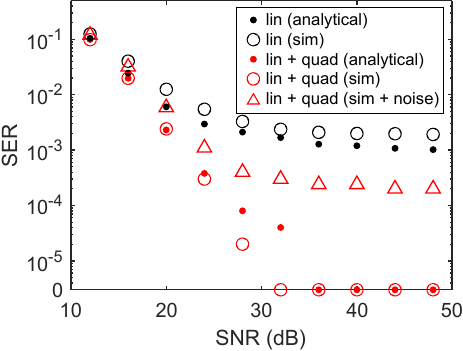}
	\caption{Comparison of the SER between a simulated experiment and a ridge regression applied to feature vectors formed by analytically computing nonlinear terms. Black markers include only linear elements and red markers indicate feature vectors with linear and quadratic contributions. Results for the analytical approach are shown as dots, and the experimental simulations are marked by circles and triangles.}
	\label{fig:ser_vs_snr_sim}
\end{figure}

We first tested the NGRC using the analytical form of $\mathbf{v}_{\text{out},n}$ in \cref{eq:vout_analytic} with $p=1$ or $p=2$, $m=9$ (extending from $n=-6\rightarrow+2$), $N_{\text{train}} = 10^4$ points, and $N_{\text{test}} = 5 \times 10^4$ points. The SER vs. SNR is shown in \Cref{fig:ser_vs_snr_sim} as solid black dots ($p=1$)  and red dots ($p=2$). Note that the baseline SER of the distorted data, $d_n$, is $\sim0.07$ at an SNR of 40 dB. As expected, adding a quadratic nonlinearity significantly reduced the SER by better approximating the inverse of \cref{eq:dn}. We also note that we observed zero errors for SNR of 35 dB or higher using a feature vector with $p=2$, confirming that the quadratic, element-wise nonlinearity combined with a memory of length $m=9$ is sufficient to compensate for the distortion in this channel. This result showed that the analytical study in \crefrange{eq:inv}{eq:WX} provides a useful framework for designing the NGRC feature vector, at least in applications where the forward model (i.e., the nonlinear distortion process) is well understood. This interpretable approach contrasts with conventional RCs, for which the hyperparameters (e.g., the number of neurons, the degree of connectedness, the form of nonlinearity) are more abstract and often require extensive empirical optimization to accomplish a given task.

After confirming a suitable form for the feature vector through our analytical investigation, we modelled the expected performance of the experimental photonic FM-NGRC (shown \cref{fig:experiment}) designed to produce this vector. In the simulation, we assumed data was encoded at a rate $f_d = 5$ GHz. To model the linear portion of the feature vector, we calculated the response of an MZI-based EOM biased at quadrature and driven with a peak voltage of $V_{\text{max}} = 0.15 V_\pi$. The quadratic contribution to the feature vector was calculated for an MZI-based EOM biased at the minimum transmission point and driven with a peak voltage of $V_{\text{max}} = 0.45 V_\pi$ (higher drive voltage was used to generate an optical signal with comparable intensity to the linear EOM). The frequency comb was modelled by simulating the response of an MZI-based EOM and phase-only EOM in series that are both driven at a frequency $f_c = 15$ GHz with drive amplitudes of $V_{\text{max}} = V_{\pi}$ and $V_{\text{max}} = 2.75 V_{\pi}$, respectively. The resulting comb comprised 9 teeth separated by $15$ GHz with $<$10 dB variation in power, setting the memory $m=9$. Delays were introduced between comb teeth by adding a quadratic phase $e^{i D f^2}$ with dispersion parameter $D = \pi/(f_d f_c)$. Finally, the waveshaper was simulated by applying a rectangular filter with  $10$ GHz bandwidth centered at each comb line.

The SER vs SNR for the simulated FM-NGRC is also shown in \Cref{fig:ser_vs_snr_sim} as black open circles ($p=1$) and red open circles ($p=2$). Again, we found that the addition of quadratic nonlinearity significantly improved on the results found for a feature vector containing only linear elements. We also found that the simulated experimental system slightly out-performed the analytical NGRC for both $p=1$ and $p=2$ and reached an error rate of 0 at 32 dB SNR for $p=2$. This result can be attributed to the fact that the response function of the MZI-based EOMs includes higher-order nonlinear terms in the nominally linear or quadratic elements. Of course, an experimental FM-NGRC will also introduce some noise which could mitigate this advantage. To model this, we added Gaussian random noise to the optical intensity in each comb tooth in the simulated experiment. After adding noise, we obtained the SER values shown with red triangles in \Cref{fig:ser_vs_snr_sim} (for a feature vector with $p=2$ and $m=9$). The SER with experimental noise plateaued at $\sim2\times10^{-4}$ for SNR above 30 dB, indicating that experimental noise may limit the performance of the photonic FM-NGRC.  

\subsection{Experimental demonstration with digital weights}
\label{sec:digital}

The results of \cref{sec:sim} indicate that the addition of quadratic elements in the feature vector drastically improves SER in idealized simulations. This section provides experimental verification of this improvement by performing tests in which the experimental setup generates the feature vectors $\mathbf{v}_{\text{out},n}$ and weights $W$ are applied in digital post-processing. In addition to feature vectors containing only linear terms ($p=1$) and vectors with linear and quadratic terms ($p=2$),  we also performed experiments in which we optimized the nonlinear encoding by sweeping through the full range of modulator bias points and chose the configuration that yielded the lowest SER. We found that this optimization can help mitigate the noise-induced performance degradation in the experimental system.

\begin{figure}[!h]
	\centering
	\includegraphics[scale=1.0]{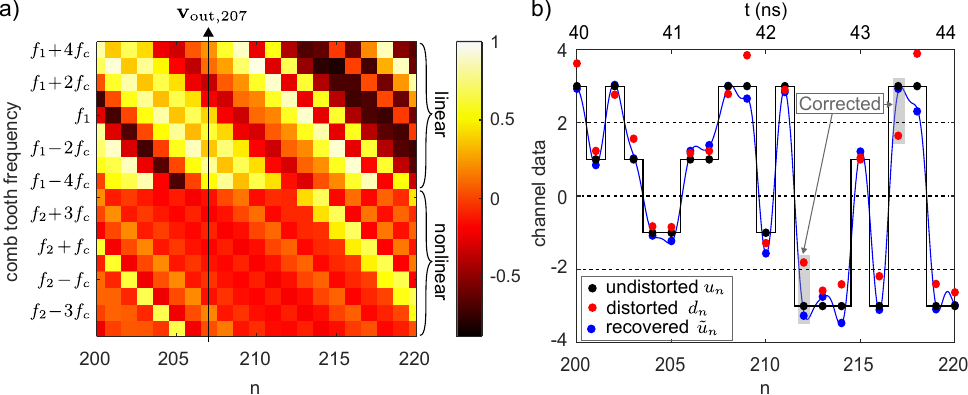}
	\caption{a) Experimentally-measured feature vectors $\mathbf{v}_{\text{out},n}$ for $n = 200$ -- $220$, with each row scaled to have unit amplitude. b) A plot of the undistorted data $u_n$, the distorted data $d_n$, and the FM-NGRC approximation of the equalized data $\tilde{u}_n$. Dotted lines mark the thresholds for mapping values to the original space $\{-3,-1,1,3\}\}$.}
	\label{fig:raw_data}
\end{figure}

The experimentally-recorded feature vectors generated by the FM-NGRC are shown in \Cref{fig:raw_data}a from $n = 200$ to 220. Each row represents the amplitude of an individual comb tooth over time, with the top 9 rows showing the first frequency comb centered around $f_1$ (carrying the linearly encoded data) and the bottom 9 rows showing the comb centered around $f_2$ (carrying the nonlinearly encoded data). Each row has been normalized for easier visualization (in practice, the output weights compensate for variations in the amplitude of each comb tooth and the efficiency of the linear and nonlinear encoding). As expected, the amplitude of each frequency comb tooth is shifted by a single time step. The feature vector at time step $n$ corresponds to a column in \Cref{fig:raw_data}a and contains 9 delayed replicas of the data after linear encoding and 9 delayed replicas of the input data after nonlinear encoding. 

To test the FM-NGRC, we recorded these feature vectors for $N_{\text{train}} = 10^4$ and $N_{\text{test}} = 10^4$ samples by measuring a single comb tooth at a time. Thus, the entire $2\times10^4$-sample training and test sequence was sent through the FM-NGRC 18 times while the waveshaper was used to select a single frequency comb tooth for digitization. We then used ridge regression to find the output weights $W$. The recovered data, $\widetilde{u}_n$, was obtained by  applying these output weights to the feature vector recorded during the testing sequence. The SER was then calculated to quantitatively evaluate the FM-NGRC performance. In practice, we temporally aligned the measured feature vector (sampled experimentally at 100 GS/s) to the 5 GS/s data rate by finding the initial sampling point that minimized the SER. This alignment also allowed us to optimize the memory structure by varying the distribution of forward and backward time-step samples included in the feature vector, $\mathbf{v}_{\text{out},n}$. We found that the optimal distribution corresponded to a feature vector containing elements from $n-6$ to $n+2$, which matches the dominant terms in \cref{eq:qn}. An example sequence of the original data, $u_n$, distorted data, $d_n$, and experimentally recovered data, $\widetilde{u}_n$, are shown in \Cref{fig:raw_data}b. The threshold levels for the 4 symbols are shown as horizontal dashed lines. The channel distortion process introduced several errors (e.g., at $n=212$ and $217$), which are corrected by the FM-NGRC. 

\begin{figure}[!h]
	\centering
	\includegraphics[scale=1.0]{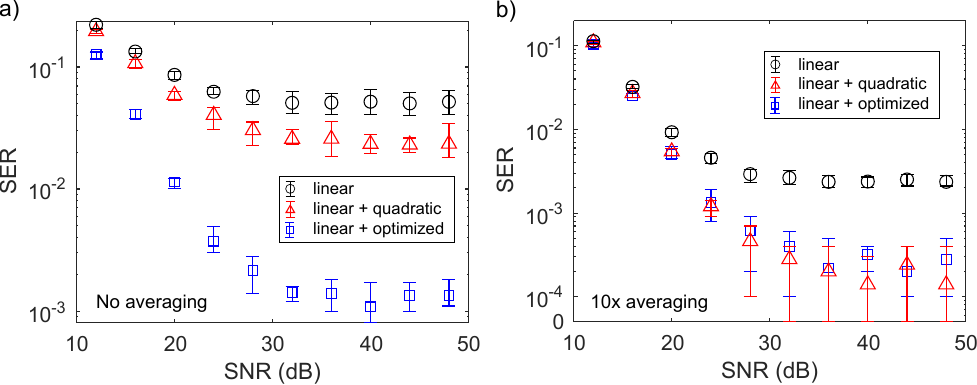}
	\caption{Experimental SER vs. SNR using digital weighting scheme. Black circles: one source with quadrature bias. Red triangles: two sources with quadrature and null bias. Blue squares: two sources with quadrature and optimized bias. Subfigure (a) includes no averaging and subfigure (b) includes a 10-run average to reduce noise effects. Error bars are computed by repeating the experiment with 5 different sets of channel data. The symbol represents the mean result and the top and bottom limits show the maximum and minimum SER for all trials.}
\label{fig:ser_vs_snr_digital_avg}
\end{figure}

\Cref{fig:ser_vs_snr_digital_avg}a shows the SER vs. SNR using experimentally-measured feature vectors. For each data point, the experiment was repeated five times with randomized channel data. The marker denotes the mean SER, and the top and bottom lines indicate the maximum and minimum values. We initially compared the SER using feature vectors including only linear elements (i.e., the 9 elements from the first optical frequency comb) to the SER obtained using feature vectors composed of linear and quadratic elements (i.e., using all 18 comb teeth recorded with the second EOM biased at the minimum transmission point). As shown in \Cref{fig:ser_vs_snr_digital_avg}a, the feature vector with quadratic elements provided lower SER, but the improvement was less pronounced than in the simulations shown in \Cref{fig:ser_vs_snr_sim}. Expecting that this discrepancy may be due to experimental noise, we averaged the measured feature vectors 10 times to reduce the noise and repeated the training and testing procedure to obtain the SER using lower-noise feature vectors. As shown in \Cref{fig:ser_vs_snr_digital_avg}b, the SER was significantly reduced, reaching $\sim2\times10^{-4}$ using the feature vector containing linear and quadratic elements. This confirmed that experimental noise was limiting the SER. Unfortunately, this averaging would preclude real-time operation. 

To mitigate the experimental noise without averaging, we varied the bias point on the second EOM to test if a bias point with higher optical transmission could reduce the noise while still providing the nonlinear transform required to equalize the channel distortion. We found an optimal bias point of $\phi=-1.41\pi$ (not far from the quadrature point at $\phi=-3\pi/2$) yielded substantially lower SER without averaging (approaching $\sim 1\times10^{-3}$). These results appear as blue squares in \Cref{fig:ser_vs_snr_digital_avg}a. This improvement was likely due to a reduction in digitization and amplified spontaneous emission (ASE) noise while including both quadratic and cubic nonlinear transformations in the feature vector. Interestingly, this optimal bias point did not substantially improve the noise after averaging, as shown in \cref{fig:ser_vs_snr_digital_avg}b; however, the SER with 10$\times$ averaging was already quite low using the linear and quadratic feature vector, reaching zero errors for several of the measurements, and longer test data sets would likely be required to resolve any meaningful differences. Nonetheless, since the optimal bias point enabled reasonably low SER without averaging, we proceeded to test the FM-NGRC using analog weights.  

\subsection{Experimental demonstration with analog weights}
\label{sec:analog}

After optimizing the FM-NGRC design, we proceeded to demonstrate continuous inference in real time using weights applied in the optical domain. As shown in \cref{fig:experiment}, the waveshaper weighted each comb tooth and directed the positive weighted comb teeth to port 1 and the negative weighted comb teeth to port 2. Since the two detectors comprising the balanced detector had slightly different responsivity (despite using nominally identical PDs), we repeated the training procedure for each PD (that is, the entire training sequence was repeated 36 times, enabling all 18 comb teeth to be recorded on both PDs). After collecting this training data, the weights were computed slightly differently to compensate for the variations in responsivity of the two PDs. 
We first calculated a standard ridge regression on the feature vector recorded on the positive PD. This initial ridge regression was used to determine the sign of the weights that would be assigned to each comb tooth. We then constructed an updated feature vector in which the elements receiving positive weights were taken from the feature vector recorded on the positive PD and elements receiving negative weights were taken from the feature vector recorded on the negative PD and multiplied by $-1$. To allow for offsets in the reconstructed data, we appended a $+1$ and $-1$ to this new feature vector. Finally, we performed a non-negative least squares regression to this composite feature vector. This yielded a $W$ vector of nominally positive weights corresponding to the fraction of each comb tooth that should be directed to either the "positive" or "negative" output port. Once the weights were determined, they were normalized so that $\text{max}(W_i) = 1$ and programmed into the waveshaper attenuation coefficients for each comb tooth. With the weights assigned, we injected the test data points into the system. The difference in the signal measured by the two PDs yields the reconstructed input channel data, $\widetilde{u}_n$. Before calculating the SER, the recovered data was scaled and offset to relate the measured photovoltage to the digital values ranging between $-3$ and 3. 

\begin{figure}[!h]
	\centering
	\includegraphics[scale=1.0]{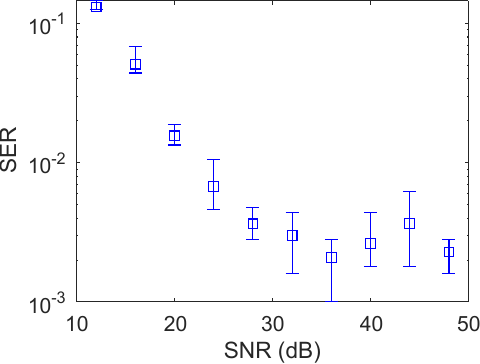}
	\caption{Experimental results for SER versus SNR for real-time channel equalization at 5 GS/s data rate with analog weighting. Squares mark the mean SER found from five different sets of randomly-generated channel data, and the top and bottom bars represent the maximum and minimum error of the five trials.}
	\label{fig:ser_vs_snr_analog}
\end{figure}

The measured SER vs. SNR using analog weights with $N_{\text{train}} = 1\times 10^4$ and $N_{\text{test}} = 5 \times 10^3$ are shown in \cref{fig:ser_vs_snr_analog}. As before, the experiment was repeated five times at each SNR level and the error bars indicate the maximum and minimum SER. We found that the SER reduces with increasing SNR until $SNR = 28$ dB, where it settles to $\sim2.5 \times 10^{-3}$. This minimum SER is attributed to the experimental noise in the current design and is comparable to the SER achieved using digital output weights without averaging shown in \cref{fig:ser_vs_snr_digital_avg}a, indicating that applying weights in the analog domain did not introduce a significant amount of additional noise. This confirmed that the FM-NGRC is capable of real-time operation at 5 GHz data rates. 

\section{Discussion} \label{sec:discussion}
This work presented a photonic FM-NGRC capable of real-time operation at GHz data rates. This scheme provides a flexible platform in which the system memory can be tailored using optical frequency combs and dispersive fiber while nonlinear elements can be introduced in the feature vector using the nonlinear response of MZI-based EOMs. To showcase the capabilities of the FM-NGRC platform, we focused on the benchmark channel equalization task, beginning with an analytical study of the feature vector form that allows for satisfactory data reconstruction. In simulations, we showed that the derived feature vector is capable of solving the channel equalization problem, achieving zero SER for SNR above 32 dB. We then experimentally constructed an FM-NGRC that generates this feature vector. Using averaging to reduce the experimental noise, the FM-NGRC achieved a $\text{SER} = 2 \times 10^{-4}$, consistent with past experimental reservoir computers \cite{duportAllopticalReservoirComputing2012,dejonckheereAllopticalReservoirComputer2014,vinckierHighperformancePhotonicReservoir2015,duportFullyAnaloguePhotonic2016,vatinExperimentalReservoirComputing2019,chemboMachineLearningBased2020}. Using a waveshaper to apply the output weights in the analog domain, we demonstrated real-time channel equalization at 5 GHz with a SER of $2\times10^{-3}$.  In the future, lower SER could be achieved by reducing the experimental noise. For example, replacing the long DCF with a chirped fiber Bragg grating would reduce the EDFA gain required to compensate for loss, thereby reducing amplified spontaneous emission noise. 

In the future, the FM-NGRC scheme could support higher data rates simply by increasing the comb spacing to encode higher bandwidth data on each comb tooth without cross-talk. The FM-NGRC could also be adapted to generate feature vectors tailored to more complex problems. For example, longer memory could be achieved by increasing the number of comb teeth and additional nonlinearities could be introduced, either as element-wise transformations using additional EOMs or as interaction terms using a modulator to introduce mixing between different comb teeth (e.g., using the final phase modulator suggested in \cref{fig:experiment}). Fortunately, programmable waveshapers are capable of simultaenously controlling hundreds of comb teeth, providing excellent scalability. The demonstration here started with distorted data in the RF domain; however, this scheme could be adapted to operate directly on data in the optical domain as well (e.g., to compensate for distortion in a fiber optics communication link). In this case, the linear encoding EOM would not be required, as the data is already in the optical domain. To achieve the element-wise nonlinear encoding imparted by the second EOM in this work, an optical-electrical-optical conversion could be deployed by detecting part of the incident signal and using the resulting photocurrent to drive an EOM at the desired bias point. 

\begin{acknowledgments}
The authors acknowledge support from the U.S. Naval Research Laboratory (6.1 Base Funding). We thank Ross Schermer for experimental assistance with RF components and frequency comb generation. We also thank George Valley for useful discussions on reservoir computing.
\end{acknowledgments}

\bibliography{FM_bib}

\end{document}